\documentclass[letterpaper, 10 pt, conference]{ieeeconf}  

\IEEEoverridecommandlockouts                              

\newtheorem{lemma}{Lemma}

\usepackage{bbm}
\usepackage{subfigure}
\usepackage{subfigmat}
\usepackage{amsfonts}
\usepackage{algorithm}
\usepackage{algpseudocode}
\usepackage{amsmath}
\usepackage{amssymb}
\usepackage{graphicx}      
\usepackage{xcolor}
\usepackage{comment}
\usepackage{subfigure}

\newcommand{\q}{q}
\newcommand{\Q}{\mathbb{Q}}
\newcommand{\X}{\mathbb{X}}
\newcommand{\Pq}{\mathsf{Pr}_\q}
\newcommand{\Su}{\underline{\mathbb{S}}}

\def\Prod#1#2{\prod\limits_{#1}^{#2}}
\def\set#1#2{\{ \; #1 \;:\;#2\;\}} 
\def\TEO#1{\textcolor{red}{TEO: #1}} 

\newcommand{\R}{\mathbb{R}}
\newcommand{\D}{\mathbb{D}}

\newcommand{\Xe}{\mathbb{X}_\varepsilon}
\newcommand{\XLT}{\underline{\X}^{j}_{LT}}
\newcommand{\wb}{\mathbf{w}}

\newtheorem{assumption}{Assumption}

\title{\LARGE \bf 
Computationally efficient stochastic MPC: a probabilistic scaling approach}

\author{Martina Mammarella$^1$, Teodoro Alamo$^2$, Fabrizio Dabbene$^{1}$, and Matthias Lorenzen$^{3}$
\thanks{$^*$This work was funded by the Italian Institute of Technology (IIT) and the Italian Ministry of Education, University and Research (MIUR) within the 2017 Projects of National Interest (PRIN 2017 N. 2017S559BB). \textit{Corresponding author}: fabrizio.dabbene@ieiit.cnr.it (Dabbene F.)}
\thanks{$^1$ Institute of Electronics, Computer and Telecommunication Engineering, National Research Council of Italy, Turin, Italy, {\tt\small martina.mammarella@ieiit.cnr.it, fabrizio.dabbene@ieiit.cnr.it}}
\thanks{$^2$ Departamento de Ingenier\'ia de Sistemas y AutomÃ¡tica, Universidad de Sevilla, Escuela Superior de Ingenieros, Camino de los Descubrimientos s/n, 41092 Sevilla, Spain, {\tt\small talamo@us.es}}
\thanks{{\tt\small lorenzen@ist.uni-stuttgart.de}}
}

\begin{document}

\maketitle
\thispagestyle{empty}
\pagestyle{empty}

\begin{abstract} In recent years, the increasing interest in Stochastic model predictive control (SMPC) schemes has highlighted the limitation arising from their inherent computational demand, which has restricted their applicability to slow-dynamics and high-performing systems. To reduce the computational burden, in this paper we extend the probabilistic scaling approach to obtain low-complexity inner approximation of chance-constrained sets. This approach provides  probabilistic guarantees at a lower computational cost than other schemes for which the sample complexity depends on the design space dimension. To design candidate simple approximating sets, which approximate the shape of the probabilistic set, we introduce two possibilities: i) fixed-complexity polytopes, and ii)  $\ell_p$-norm based sets. Once the candidate approximating set is obtained, it is scaled around its center so to enforce the expected probabilistic guarantees. The resulting scaled set is then exploited to enforce constraints in the classical SMPC framework. The computational gain obtained with the proposed approach with respect to the scenario one is demonstrated via simulations, where the objective is the control of a fixed-wing UAV performing a monitoring mission over a sloped vineyard.
\end{abstract}

\section{Introduction}
\label{sec:intro}
In recent years, the performance degradation of model predictive control (MPC) schemes in the presence of uncertainty has driven the interest towards stochastic MPC, to overcome the inherent conservativeness of robust approaches.
A probabilistic description of the disturbance or uncertainty allows to optimize the average performance or appropriate risk measures. Furthermore, allowing a (small) probability of constraint violation, by introducing so-called chance constraints, seems more appropriate in some applications.
As highlighted in \cite{farina2016stochastic}, current SMPC methods can be divided in two main groups, depending on the approach followed to solve the chance-constrained optimization problem: (i) analytic approximation methods; and (ii) randomized \cite{tempo2012randomized} and scenario-based methods.
For the analytic approximation methods, the probabilistic properties of the uncertainty are exploited to reformulate the chance constraints in a deterministic form. For the second class of methods, the craved control performance and constraint satisfaction are guaranteed properly generating a sufficient number of uncertainty realizations and on the solution of a suitable constrained optimization problem, as proposed in \cite{calafiore2006scenario}, \cite{schildbach2014scenario}. The main advantage of this class of stochastic MPC algorithms is given by the inherent flexibility to be applied to (almost) every class of systems, including any type of uncertainty and both state and input constraints, as long as the optimization problem is convex. On the other hand, they share two main drawback: i) slowness, which has limited their application to problems involving slow dynamics and where the sample time is measured in tens of seconds or minutes; and ii) a significant computational burden required for real-time implementation, narrowing the application domains to those involving low-computation assets. Some examples are \cite{grosso2017stochastic} for water networks, \cite{nasir2015randomised} for river flood control, \cite{van2006stochastic} for chemical processes, and \cite{vignali2017energy} for energy plants. 

An efficient solution to the aforementioned disadvantages was proposed in \cite{matthias1} where the SMPC controllers design is based on an \textit{offline} sampling approach and only a predefined number of necessary samples are kept for online implementation. In this approach, the sample complexity is linearly dependent to the design space dimension and the sampling procedure allows to obtain offline an inner approximation of the chance-constrained set. This approach has been extended to a more generic setup in \cite{Mammarella:18:Control:Systems:Technology} and experimentally validated for the control of a spacecraft during rendezvous maneuvers. Beside the efficacy of the approach, the results highlighted the need to further reduce the computational load and the slowness of the proposed approach to comply with faster dynamics and low-cost, low-performance hardware.

Among challenging applications, the control of unmanned aerial vehicles (UAVs) during assorted scenarios, have been triggering the attention of MPC community. These platforms are typically characterized by fast dynamics and equipped with computationally-limited autopilots. In the last decade, different receding horizon techniques have been proposed, see e.g. \cite{kamel,Alexis2016,stastny,michel}, including a stochastic approach by \cite{mammarella2018sample}. In this case, preliminary analysis have confirmed the effectiveness of the proposed offline sampling-based SMPC (OS-SMPC) strategy but the results highlighted also the need to further reduce the dimension of the optimization problem to comply with hardware requirements. 

The main contribution of this paper is to propose a new methodology that combines the probabilistic-scaling approach proposed in \cite{alamo2019safe}, which allows to obtain a low-complexity inner approximation of the chance constrained set, with the SMPC approach of \cite{matthias1,Mammarella:18:Control:Systems:Technology}. In \cite{alamo2019safe}, authors show how to scale a given set of manageable complexity around its center to obtain, with a user-defined probability, a region that is included in the chance constrained set. In this paper, we extend the aforementioned approach showing how it is possible to reduce the sample complexity via probabilistic scaling exploiting so-called simple approximating sets (SAS). The starting point consists in obtaining a first simple approximation of the ``shape'' of the probabilistic set. To design a candidate SAS, we propose two possibilities. The first one is based on the definition of an approximating set by drawing a fixed number of samples. On the other hand, the second case envisions the use of $\ell_p$-norm based sets, first proposed in \cite{dabbene2010complexity}. In particular, we consider as SAS a $\ell_1$-norm \textit{cross-polytope}  and a $\ell_\infty$-norm \textit{hyper-cube}. Solving a standard optimization problem, it is possible to obtain the center and the shape of the SAS, which will be later scaled to obtain the expected probabilistic guarantees following the approach described in \cite{alamo2019safe}. Then, the scaled SAS is used in the classical SMPC algorithm to enforce constraints.

To validate the proposed approach, an agriculture scenario has been selected, because of the increasing interest of using drones in the agriculture 4.0 framework, as explained in \cite{sylvester2018agriculture}, due to their great potential to support and address some of the most pressing dares in farming. And real-time quality data and crop monitoring are two of those challenges. In particular, UAVs could represent a favorable alternative to conventional farming machines, whenever clear advantages with respect to traditional methods, in terms of higher efficiency in operations, reduced environmental impact or enhanced human health and safety are sought. For this paper, the control objective envisions the proposed approach applied to a fixed-wing UAV performing a monitoring mission over a sloped vineyard, following a pre-defined snake path. The performance of the proposed approach in terms of tracking capabilities and computational load has been compared with those obtained exploiting the ``classical'' OS-SMPC scheme proposed in \cite{mammarella2018sample}.  

{\textit{Notation}: The set $\mathbb{N}_{>0}$ denotes the positive integers, the set $\mathbb{N}_{\geq 0} = \left\{0\right\} \cup\mathbb{N}_{>0}$ the non-negative integers, and $\mathbb{N}_a^b$ the integers interval $[a,b]$. Positive (semi)definite matrices $A$ are denoted $A\succ 0$ $(A\succeq 0)$ and $\|x\|_A^2\doteq x^TAx$. For vectors, $x\succeq 0$ ($x\preceq 0$) is intended component-wise.  $\mathsf{Pr}_a$ denotes the probabilistic distribution of a random variable $a$. Sequence of scalars/vectors are denoted with bold lower-case letters, i.e. $\textbf{v}$. }

\section{Offline Sampling-based Stochastic MPC}
In this section, we first recall the Stochastic MPC Framework proposed in \cite{matthias1,Mammarella:18:Control:Systems:Technology}. 

\subsection{Problem setup}\label{sec:setup}
We consider the case of a discrete-time system subject to generic uncertainty $w_{k} \in \mathbb{R}^{n_{w}}$
\begin{equation}
x_{k+1} = A(w_{k})x_{k}+B(w_{k})u_{k}+a_{w}(w_{k}),
\label{eq:sys}
\end{equation}
with state $x_{k} \in \mathbb{R}^{n}$, control input $u_{k} \in \mathbb{R}^{m}$, and the vector valued function $a_w(w_{k})$ represent the additive disturbance affecting the systems states. The system matrices $A(w_{k})$ and $B(w_{k})$, of appropriate dimensions, are (possibly nonlinear) functions of the uncertainty $w_{k}$ at step $k$. 
The disturbances $(w_{k})_{k\in \mathbb{N}_{\geq 0}}$ are modeled as realizations of the stochastic process $(W_{k})_{k\in \mathbb{N}_{\geq 0}}$, on which take the following assumptions.

\begin{assumption}[Random Disturbances] \label{bound_rand_dist}
The disturbances $W_{k}$, for $k\in \mathbb{N}_{\geq 0}$, are independent and identically distributed (i.i.d.), zero-mean random variables with support $\mathbb{W}\subseteq \mathbb{R}^{n_{w}}$.
Moreover, let $\mathbb{G}=\left\{(A(w_{k}),B(w_{k}),a_{w}(w_{k})\right\}_{w_{k}\in \mathbb{W}}$, a polytopic outer approximation with $N_c$ vertexes $\bar{\mathbb{G}}\doteq co\left\{A^{j},B^{j},a_{w}^{j}\right\}_{j\in \mathbb{N}_{1}^{N_{c}}}\supseteq \mathbb{G}$ exists and is known.
\end{assumption}

\vskip 3mm
We can notice that the system can be augmented by a filter to model a specific stochastic processes of interest. The assumption of independent random variables in necessary to perform the offline computations discussed next while the need of a known outer bound is required to establish a safe operating region (see~\cite{matthias1} for details). We remark that the system's representation in \eqref{eq:sys} is very general, and encompasses e.g. those in \cite{matthias1,Mammarella:18:Control:Systems:Technology,matthias2}. Given the model~\eqref{eq:sys} and a realization of the state $x_k$ at time $k$, state predictions $l$ steps ahead are random variables, as well and are denoted $x_{l|k}$, to differentiate it from the realization $x_{l+k}$. Similarly $u_{l|k}$ denotes predicted inputs that are computed based on the realization of the state $x_k$.

The system is subject to $p$ state and input chance constraints of the form\footnote{%
The case where one wants to impose \textit{hard} input constraints can be also be formulated in a similar framework, see e.g.~\cite{matthias1}.}
\begin{align}
 &&\mathsf{Pr}_\wb \left\{  [H_x]_j^T x_{l|k} + [H_u]_j^T u_{l|k} \le 1 |x_k \right\}\ge 1-\varepsilon_j,\nonumber\\
 && \qquad l \in \mathbb{N}_{>0}, \, j \in \mathbb{N}_1^p, \label{eq:origConstr} 
\end{align}
with $\varepsilon_j \in (0,1)$, and
$H_x\in \R^{p \times n}$, $H_u \in \R^{p \times m}$, where $[H]_j^T$ denotes the $j$-th row of matrix $H$. The probability $\mathsf{Pr}_\wb$ is measured with respect to the sequence $\wb=\{w_i\}_{i>k}$. Hence, equation~\eqref{eq:origConstr} states that the probability of violating the linear constraint $[H_x]_j^T x + [H_u]_j^T u \le 1$ for any future realization of the disturbance should not be larger than $\varepsilon_j$.

The objective is to derive an asymptotically stabilizing control law for the system~\eqref{eq:sys} such that, in closed loop, the constraints~\eqref{eq:origConstr} are satisfied.

\subsection{Stochastic Model Predictive Control}
To solve the constrained control problem, a stochastic MPC algorithm is considered. The approach is based on repeatedly solving a stochastic optimal control problem over a finite, moving horizon, but implementing only the first control action. 
Defined the control sequence as $\mathbf{u}_{k} = (u_{0|k}, u_{1|k}, ..., u_{T-1|k})$, the prototype optimal control problem that is to be solved at each sampling time is given minimizing the cost function
\begin{multline}
      J_T(x_k,\mathbf{u}_{k}) =\\
    \mathbb{E}\left\{ \sum_{l=0}^{T-1} \left( x_{l|k}^\top Q x_{l|k} + u_{l|k}^\top Ru_{l|k} \right) + x_{T|k}^\top P x_{T|k} ~|~ x_k\right\}
  \label{eq:origCostFnc}
\end{multline}
with $Q\succ 0$, $R \succ 0$, and appropriately chosen $P \succ 0$, subject to the system dynamics~\eqref{eq:sys} and constraints~\eqref{eq:origConstr}.

The online solution of the stochastic MPC problem remains a challenging task but several special cases, which can be evaluated exactly, as well as methods to approximate the general solution have been proposed in the literature. The approach followed in this work was first proposed in~\cite{matthias1,Mammarella:18:Control:Systems:Technology}, where an offline sampling scheme was introduced. Therein, with a prestabilizing input parametrization
\begin{equation}
u_{l|k}=Kx_{l|k}+v_{l|k},
\label{eq:prestabilizingInput}
\end{equation}
with suitably chosen control gain $K\in\mathbb{R}^{n\times m}$ and free optimization variables $v_{l|k} \in \mathbb{R}^m$, equation~\eqref{eq:sys} is solved explicitly for the predicted states $x_{1|k},\ldots,x_{T|k}$ and predicted inputs $u_{0|k},\ldots,u_{T-1|k}$.
In this case, the expected value of the finite-horizon cost~\eqref{eq:origCostFnc} can be evaluated \textit{offline}, leading to a quadratic cost function of the form
\begin{equation}
J_{T}(x_{k},\mathbf{v}_{k})=[x_{k}^{T}\,\, \mathbf{v}_{k}^{T} \,\,\textbf{1}_{n}^{T}]\tilde{S}\begin{bmatrix}
        x_{k} \\
        \textbf{v}_{k} \\
        \textbf{1}_{n}\\
        \end{bmatrix}
        \label{eq:cost_new_1}
\end{equation}
in the deterministic variables $\mathbf{v}_{k} = (v_{0|k}, v_{1|k}, ..., v_{T-1|k})$ and $x_{k}$.
The reader can refer to \cite[Appendix A]{Mammarella:18:Control:Systems:Technology} for a detailed derivation of the cost matrix $\tilde{S}$.

Focusing now on the constraint definition, we can notice that by introducing the uncertainty sequence $\wb_k=\{w_l\}_{l=k,...,k+T-1}$, we can rewrite the $j$-th chance constraint defined by equation~\eqref{eq:origConstr} as 
\begin{multline}
    \Xe^j= 
    \left\{\begin{bmatrix}
    x_k\\
    \mathbf{v}_k
    \end{bmatrix}\in\mathbb{R}^{n+mT} ~|~ \right. \\
    \left. \mathsf{Pr}_{\wb_k}\left\{f_j^T(\wb_k)
    \begin{bmatrix}
    x_k\\
    \mathbf{v}_k
    \end{bmatrix} \leq 1 \right\} \geq 1-\varepsilon \right\},
    \label{eq:Xej}
\end{multline}
with $f_j$ being a function of the  sequence of random variables $\wb_k$. Again, the reader is referred to \cite{Mammarella:18:Control:Systems:Technology} for details on the derivation of $f_j$. The results in \cite{matthias1} show that, by exploiting results from statistical learning theory (cf. \cite{vidyasagar,alamo2009randomized}),
we can construct an inner approximation $\underline{\X}^{j}$ of the constraint set $\Xe^j$ by extracting $N_{LT}$ i.i.d. samples $\wb_k^{(i)}$ of $\wb_k$ and taking the intersection of the sampled constraints, i.e.
\begin{multline}
  \XLT=
      \left\{\begin{bmatrix}
    x_k\\
    \mathbf{v}_k
    \end{bmatrix}\in\mathbb{R}^{n+mT} ~|~ \right. \\
    \left. 
  f_j^T(\wb_k^{(i)})\begin{bmatrix}
    x_k\\
    \mathbf{v}_k
    \end{bmatrix}\leq 1,\;i=1,\ldots,N_{LT}\right\},
\end{multline}
In particular, it has been shown in \cite{matthias1}, that for given probabilistic levels $\delta\in(0,1)$ and $\varepsilon_j\in(0,0.14)$, choosing the sample complexity
\begin{multline}
\label{eq:Ntilde}
N_{LT}^j \ge \tilde{N}(n+mT,\varepsilon_j,\delta) \\
\doteq \frac{4.1}{\varepsilon}\Big(\ln \frac{21.64}{\delta}+4.39(n+mT)\,\log_{2}\Big(\frac{8e}{\varepsilon_j}\Big)\Big),
\end{multline}
guarantees that with probability at least $\delta$ the sample approximation $\XLT$ is included in the original chance constraint $\Xe^j$, i.e. 
\begin{equation}
    \mathsf{Pr}\left\{\XLT\subseteq \Xe^j\right\}\geq 1-\delta, \quad j=1,.., p.
\end{equation}
Hence, exploiting these results, we obtain that the stochastic MPC problem can be well approximated by the following linearly constrained quadratic program
\begin{align}
\min_{\mathbf{v}_k} ~&J_T(x_k \mathbf{v}_k) \\
\text{s.t. } & (x_k, \mathbf{v}_k) \in \XLT, \quad j=1,.., p
\end{align}

While the result reduces the original stochastic optimization program to an efficiently solvable quadratic program, the ensuing number of constraints, equal to
\[
N_{LT}=\sum_{i=1}^T N_{LT}^j,
\]
may still be too large. 
For instance, even for a moderately sized MPC problem with $n=5$ states, $m=2$ inputs and horizon of $T=10$, and for a reasonable choice of probabilistic $\varepsilon^j=0.05$, $\delta=10^{-6}$, we get $N_{LT}^j=20,604$. For this reason, in \cite{matthias1} a  post-processing analysis of the constraint set was proposed for removing redundant constraints. While it is indeed true that all the cumbersome computations may be performed offline, it is still the case that in applications with stringent requirements on the solution time the final number of inequalities may easily become unbearable. This observation motivates the approach presented in the next section, which builds upon the results presented in~\cite{alamo2019safe}, showing how the probabilistic scaling approach leads to approximations of ``controllable size," that can be directly used in applications.

\section{Complexity Reduction via Probabilistic Scaling}
\label{sec:scaling}
In this section, we consider the very general problem of finding a decision variable vector $\xi$, restricted to a set $\Xi\subseteq\R^{n_\xi}$, subject $p$ uncertain linear inequalities. Formally, we consider uncertain inequalities
of the form
\begin{equation}\label{ineq:F:g}
F(q)\xi \le g(\q)
\end{equation}
where $F(\q)\in\mathbb{R}^{p\times{n_\xi}}$ 
and $g(q)\in\mathbb{R}^{n_p}$ are continuous function of the uncertainty vector $q\in\mathbb{R}^{n_q}$. 
The uncertainty vector $q$ is assumed to be of random nature, with given probability distribution $\Pq$ and (possibly unbounded) support $\Q$. Hence, to each sample of $q$ corresponds a different set of linear inequalities.
We aim at finding an approximation of the $\varepsilon$-chance-constraint set, defined as
\begin{equation}
    \label{XE}
\Xe \doteq
\Bigl\{\xi\in\Xi\;|\;
\mathsf{Pr}_q\left\{
F(q)\xi \le g(q)
\right\}\ge 1-\varepsilon
\Bigr\}
\end{equation}
that represents the region of the design space $\Xi$ for which this probabilistic constraint is satisfied. Note that this captures exactly the SMPC setup discussed in the previous section. Indeed, the chance-constrained set in \eqref{eq:Xej} is a special instance of \eqref{XE}, with $\xi=[x_k^T\quad\mathbf{v}_k^T]^T$ and $q=\wb_k$.
\begin{figure}[!ht]
\centering
\includegraphics[trim= 20mm 50mm 20mm 50mm, clip=true,width=.49\columnwidth]{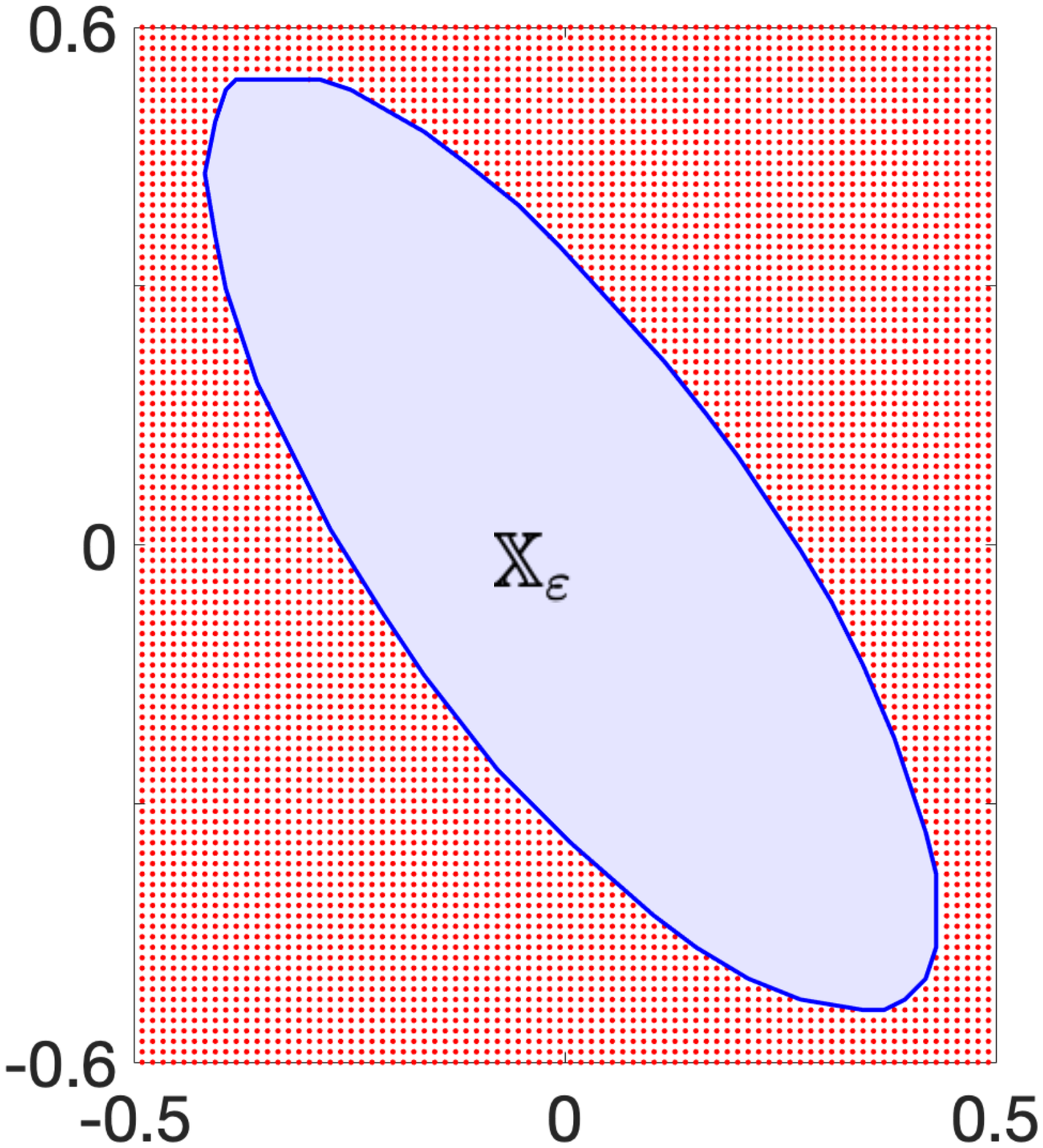}%
\label{f:Xe2D1}
\caption{Example of chance-constraint set for $\varepsilon=0.05$, for 2D scalar linear constraints of the form $f(q)^T\xi\le 1$, with $f(q)=q_1 \cdot q_2 \in \R^2$,  $q_1\in\R$ uniformly distributed in the interval $[0.5,1.5]$ and $q_2\in\R^2$ Gaussian with  variance $\Sigma$. The set was obtained by evaluating the empirical probability via random sampling.}
\end{figure}

The characterization of the chance constrained set has several application in robust and stochastic control. A classical approach is to find inner convex approximation of the probabilistic set $\Xe$, obtained for instance by means of applications of Chebyshev-like inequalities, see e.g. \cite{yan2018stochastic} and \cite{hewing2018stochastic}. A recent approach, which is the one applied in the previous section to the SMPC problem, is instead based on the derivation of probabilistic approximations of the chance constraints set $\Xe$ through sampling of the uncertainty. That is, we aim at constructing a set $\underline{\X}$ which is contained in $\Xe$ \textit{with high probability}.

Denote $F_j(q)$ and $g_j$ the $j$-th row of $F(q)$ and $j$-th component of $q$ respectively. Consider the binary functions 
$$ h_j(\xi,q) \doteq \left\{ \begin{array}{rl} 0 & \mbox{ if } F_j(q)\xi \le g_j(q) \\ 1 & \mbox{ otherwise}  \end{array}\right.,\; j=1,\ldots, p. $$ 
Now, if we define $$ h(\xi,q) \doteq  \Prod{j=1}{p} h_j(\xi,q).$$
we have that $h$ is an $(1,p)$-boolean function since it can be expressed as a function of $p$ boolean functions, each of them involving a polynomial of degree 1. See e.g. \cite[Definition 7]{alamo2009randomized} for a precise definition of this sort of boolean functions. Suppose that we draw $N$ i.i.d. samples $q^{(i)}$, $i=1,\ldots,N$. Then, we can consider the (empirical) region $\X_N$ defined as
$$ \X_N\doteq \set{\xi \in \R^{n_\xi}}{h(\xi,q^{(i)})=0, \, i=1,\ldots,N}.$$ It has been proved in \cite[Theorem 8]{alamo2009randomized}, that if $\epsilon\in(0,0.14)$ and $N$ is chosen such that\footnote{%
Note the difference under the $\log_2$ with respect to  \eqref{eq:Ntilde}.}
 $$ N \geq  \frac{4.1}{\epsilon} \left( \ln\frac{21.64}{\delta} + 4.39 n_\xi \log_2 \left(  \frac{8ep}{\epsilon}\right)   \right)$$
then $\X_N \subseteq \Xe$ with a probability no smaller than $1-\delta$.

 We notice that $\X_N$ is a convex set, which is a desirable property in an optimization framework. However, the number of required samples $N$ might be prohibitive for a real-time application. To tackle this issue, in this paper we exploit an appealing alternative approach proposed in \cite{alamo2019safe}, and we specialize it to the problem at hand.
This work proposes a probabilistic scaling approach to obtain, with given confidence, an inner approximation of the chance constrained set $\Xe$ avoiding the computational burden due to the sample complexity raising in other strategies.

The main idea behind this approach consist in first obtaining a simple initial approximation of the ``shape" of the probabilistic set $\Xe$ by exploiting simple approximating sets of the form  
\[
x_c\oplus\Su.
\]
This set is not required to have \textit{any} guarantees of probabilistic nature.
Instead, to derive such probabilistic guaranteed set, a scaling procedure  is devised. In particular, an optimal scaling factor $\gamma$ is derived so that the set scaled around its center $x_c$
\begin{equation}
\label{SASgamma}
 \Su(\gamma)\doteq x_c\oplus\gamma \Su.   
\end{equation}
is guaranteed to be an inner approximation of $\Xe$ with the desired confidence level $\delta$.

\subsection{Simple Approximating Sets}
The idea at the basis of the proposed approach is to define Simple Approximating Sets (SAS), which represent specifically defined sets with a low -- and pre-defined -- number of constraints.
First, we note that the most straightforward way to design a candidate SAS is to draw a fixed number $N_S$ of uncertainty samples, and to construct a sampled approximation as follows:\\
\noindent\textbf{1. Sampled-poly}
\begin{equation}
    \Su_S=\bigcap_{i=1}^{N_S}\X_i
\end{equation}
where
\begin{equation}
\label{Xi}
\X_i\doteq\Bigl\{\xi\in\Xi\;|\;
F(q^{(i)})\le g(q^{(i)}),
\quad i=1,\ldots,N_{S}
\Bigr\}
\end{equation}
Clearly, if $N_S<<N_{LT}$, the probabilistic properties of $\Su_S$ before scaling will be very bad. However, at this point we do not care, since the probabilistic scaling proposed in Section \ref{sec:sas-scale} will take care of this.

A second way to construct a SAS considered in this paper exploits a class of $\ell_p$-norm based sets introduced in \cite{dabbene2010complexity} as follows 
\begin{equation}
    \mathcal{A}(x_c,P) \doteq \left\{\xi \in \mathbb{R}^{n_{\xi}}\;|\;\xi=x_c+Pz,z\in\mathcal{B}_p\right\},
    \label{eq:gener_xc}
\end{equation}
where $\mathcal{B}_p \subset \mathbb{R}^{n_{\xi}}$ is the unit ball in the \textit{p} norm, $x_c$ is the center and $P=P^T\succeq 0$ is the so-called \textit{shape} matrix. In particular, we note that for $p=1,\infty$ these sets take the form of polytopes with fixed number of facets/vertices. Hence, we introduce the following two SAS:\\
\noindent \textbf{2. $\ell_1$-poly}
    \begin{equation}
        \Su_1=\left\{\xi\in\mathbb{R}^{n_{\xi}}\;|\;\xi=x_c+Pz,\;\|z\|_1\leq 1 \right\},
    \end{equation}
    defined starting from a \textit{cross-polytope}, also known as \textit{diamond}, of order $n_{\xi}$ with $2n_{\xi}$ vertices and $2^{n_{\xi}}$ facets.\\
    
\noindent \textbf{3. $\ell_{\infty}$-poly} 
        \begin{equation}
        \Su_{\infty}=\left\{\xi\in\mathbb{R}^{n_{\xi}}\;|\;\xi=x_c+Pz,\;\|z\|_{\infty}\leq 1 \right\},
    \end{equation}
        defined starting from a \textit{hyper-cube} of dimension $n_{\xi}$ with $2^{n_{\xi}}$ vertices and $2n_{\xi}$ facets.\\

Hence, the problem becomes designing the center and shape parameters $(x_c,P)$ of the set $\Su_1$ (resp.\ $\Su_\infty$) so that they represent in the best possible way the set $\Xe$. To this end, we start from a \textit{sampled design polytope} 
\[
\D=
\bigcap_{i=1}^{N_D}\X_i,
\]
with a fixed number of samples $N_D$, and construct the largest set $\Su_1$ (resp.\ $\Su_\infty$) contained in $\D$. It is easily observed that to obtain the largest $\ell_1$-poly inscribed in $\D$, we need to solve the following convex optimization problem
\begin{align}
\label{eq:test}
\max\limits_{x_c,C} \,\,\,\,& \text{tr}(P)\\
 \text{s.t.}&\quad  P\succeq 0,\nonumber\\
&\quad  f_i^TPz^{[j]}\leq g_i-f_i^Tx_c\nonumber\\
&\quad \quad  i=1,\ldots,N_D, \quad z^{[j]}\in\mathcal{V}_1,
     \end{align}
where $\mathcal{V}_1 =\{z^{[1]},\ldots,z^{[2n_\xi]}\}$ are the vertices of the unit cross-polytope while the vertices of the optimal  $\ell_1$-poly can then be obtained as
\begin{equation}
    \xi^{[j]}=x_c+Pz^{[j]}, \quad  j=1,\ldots,2n_\xi.
\end{equation}
It should be remarked that, from these vertices, one could then recover the corresponding $2^{n_\xi}$ linear inequalities, each one defining a facet of the rotated diamond. However, this procedure, besides being computationally extremely demanding (going from a vertex-description to a linear inequality description of a polytope is known to be NP hard, \cite{kaibel2003some}), would lead to an exponential number of linear inequalities, thus rendering the whole approach not viable. Instead, we exploit the following equivalent formulation of \eqref{eq:gener_xc}, see e.g.~\cite{dabbene2010complexity} for details
\begin{equation}
    \Su_1=\left\{\xi\in\mathbb{R}^{n_\xi}\;|\;\|Mx-c\|_1\leq 1\right\}
\end{equation}
where $M\doteq P^{-1}$ and $c\doteq P^{-1}x_c$. From a computational viewpoint, this second approach results to be more appealing. Indeed, using a slack variable $\zeta$, it is possible to obtain the following system of $3n_\xi+1$ linear inequalities
\[
\left\{
\begin{array}{ll}
    m_i^T\xi-c_i \leq \zeta_i,& i=1,\ldots,n_\xi\\
    -m_i^T\xi+c_i \leq \zeta_i,& i=1,\ldots,n_\xi\\
    \zeta_i\geq 0,& i=1,\ldots,n_\xi\\
    \sum_i^{n_\xi}\zeta_i\leq 1, 
\end{array}
\right.
\]
%
%
The same convex optimization problem of \eqref{eq:test} could be solved to define the center and the shape of the \textit{largest} $\ell_{\infty}$-poly inscribed in $\D$. However, this would involve an exponential number of vertices $2^{n_\xi}$. To avoid this, an approach based on Farkas lemma can be adopted, exploiting again a formulation in terms of linear inequalities. The details are not reported here due to space limitations. In this second case, obtained the center $x_c$ and the rotation matrix $P$, the corresponding $\mathcal{H}$-poly has only $2n_\xi$ hyper-planes, each one representing a different linear inequality.

Once the initial SAS, $\Su_S$ and the $\ell_1$- and $\ell_{\infty}$-polys, i.e. $\Su_1$ and $\Su_{\infty}$ respectively, has been evaluated in terms of linear inequalities, the probabilistic scaling approach can be applied to determine the corresponding scaling factor~$\gamma$.
The scaling procedure is described in details in the paper \cite{alamo2019safe}. For the sake of completeness, in the next subsection we recall its basic ideas and illustrate its application to the SAS case.

\subsection{SAS probabilistic scaling}
\label{sec:sas-scale}
Given a candidate SAS set, the following simple algorithm can be used to guarantee with prescribed probability $1-\delta$ that the scaled set $\Su(\gamma)$ is a good inner approximation of~$\Xe$.

\begin{algorithm}
\caption{Probabilistic SAS Scaling}
\begin{algorithmic}[1]
\State Given probability levels $\varepsilon$ and $\delta$, let
\[    
N_\gamma \ge  \frac{7.67}{\varepsilon} \ln\frac{1}{\delta}\text{ and }
r=\left\lceil \frac{\varepsilon N_\gamma}{2}\right\rceil.
\] 
\State Draw $N_\gamma$ samples of the uncertainty 
$q^{(1)},\ldots,q^{(N_\gamma)}$
\For  {$i=1$ to $N_\gamma$}
\State
Solve the optimization problem
\begin{align}
    \gamma_i \doteq &\arg\max \gamma \\
    &\text{s.t.}\quad  \Su(\gamma) \subseteq \X_i \nonumber
\end{align}
\EndFor
\State
Return the $r$-th smallest value of
 $\gamma_i$.

\end{algorithmic}
\end{algorithm}

A few comments are at hand regarding the algorithm above. In step 4, for each uncertainty sample $q^{(i)}$ one has to solve a convex optimization problem, which amounts at finding the largest value of $\gamma$ such that $\Su(\gamma)$ is contained in the set $\X_i$ defined in \eqref{Xi}. Then, in step 6, one has to reorder the set $\{ \gamma_1, \gamma_2, \ldots, \gamma_{N_\gamma}\}$ so that the first element is the smallest one, the second element is the second smallest one, and so on and so fort, and then return the $r$-th element of the reordered sequence. The following Lemma applies to Algorithm~1.
\vskip 0.3cm
\begin{lemma}
Given a candidate SAS set in the form $\Su(\gamma)= x_c\oplus\gamma \Su$, assume that $x_c\in\Xe$. Then, Algorithm~1 guarantees that 
\[
\Su(\gamma)\subseteq \Xe
\]
with probability at least $1-\delta$.
\end{lemma}
\vskip 0.3cm
Proof to Lemma 1 is reported in Appendix. 

\begin{figure}[!ht]
\centering
\includegraphics[trim= 20mm 5mm 20mm 5mm, clip=true,width=.49\columnwidth]{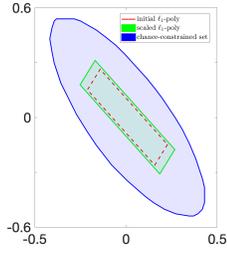}%
\label{f:Xe2D}
\caption{Visualization of the probabilistic scaling procedure for the 2D chance-constraint set considered in Figure~\ref{f:Xe2D}. 
The initial set (in red) was obtained  by constructing the largest  $\Su_1$-poly enclosed in the set $\Su$.}
\end{figure}

\subsection{Illustrating Example}
\begin{figure}[!ht]
\centering
\subfigure[$\gamma_1=0.7183$]{\includegraphics[trim= .5cm .5cm .5cm 1cm, clip=true,width=.49\columnwidth]{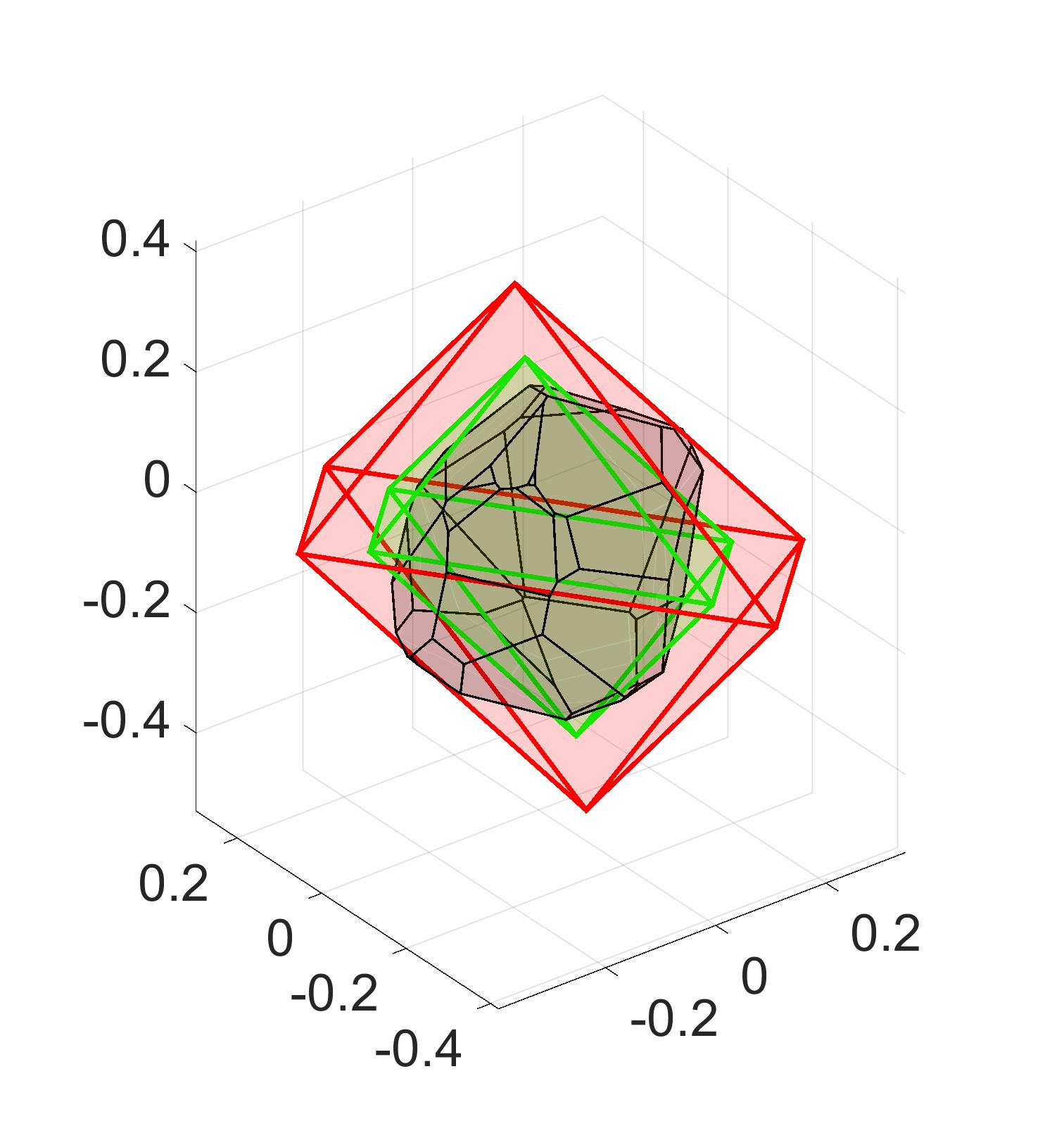}%
\label{f:diam_N100_16JAN2020}}
\hfil
\subfigure[$\gamma_\infty=0.6527$]{\includegraphics[trim= .5cm .5cm .5cm 2cm, clip=true,width=.49\columnwidth]{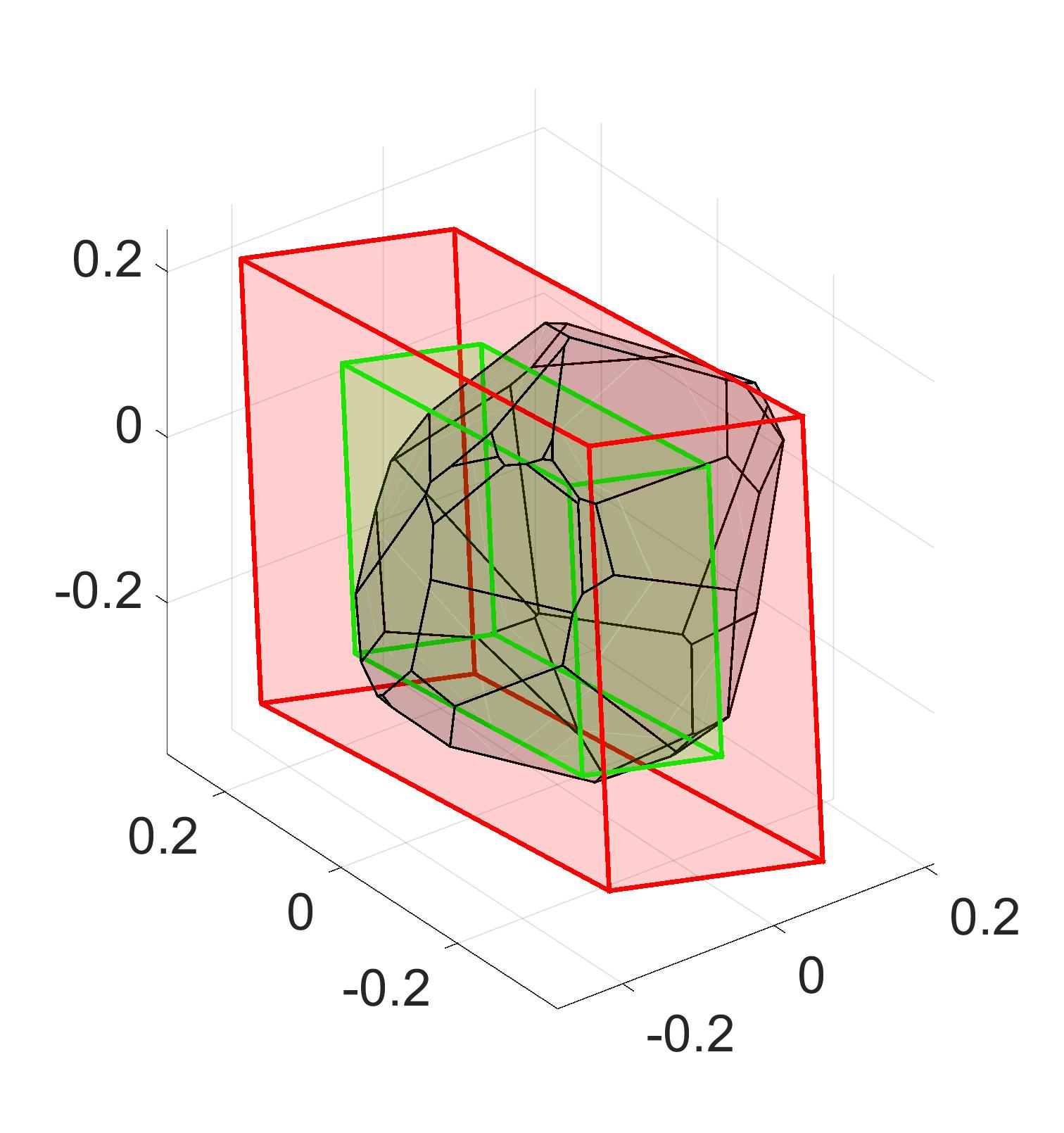}%
\label{f:hyper_N100_16JAN2020}}
\caption{Scaled $\ell_{1}$-poly (a) and $\ell_\infty$-poly (b) obtained starting from $N_D=100$.}
\label{f:diam_hyper_100}
\end{figure}

\begin{figure}[!ht]
\centering
\subfigure[$\gamma_1=1.1509$]{\includegraphics[trim= .5cm .5cm .5cm 2cm, clip=true,width=.49\columnwidth]{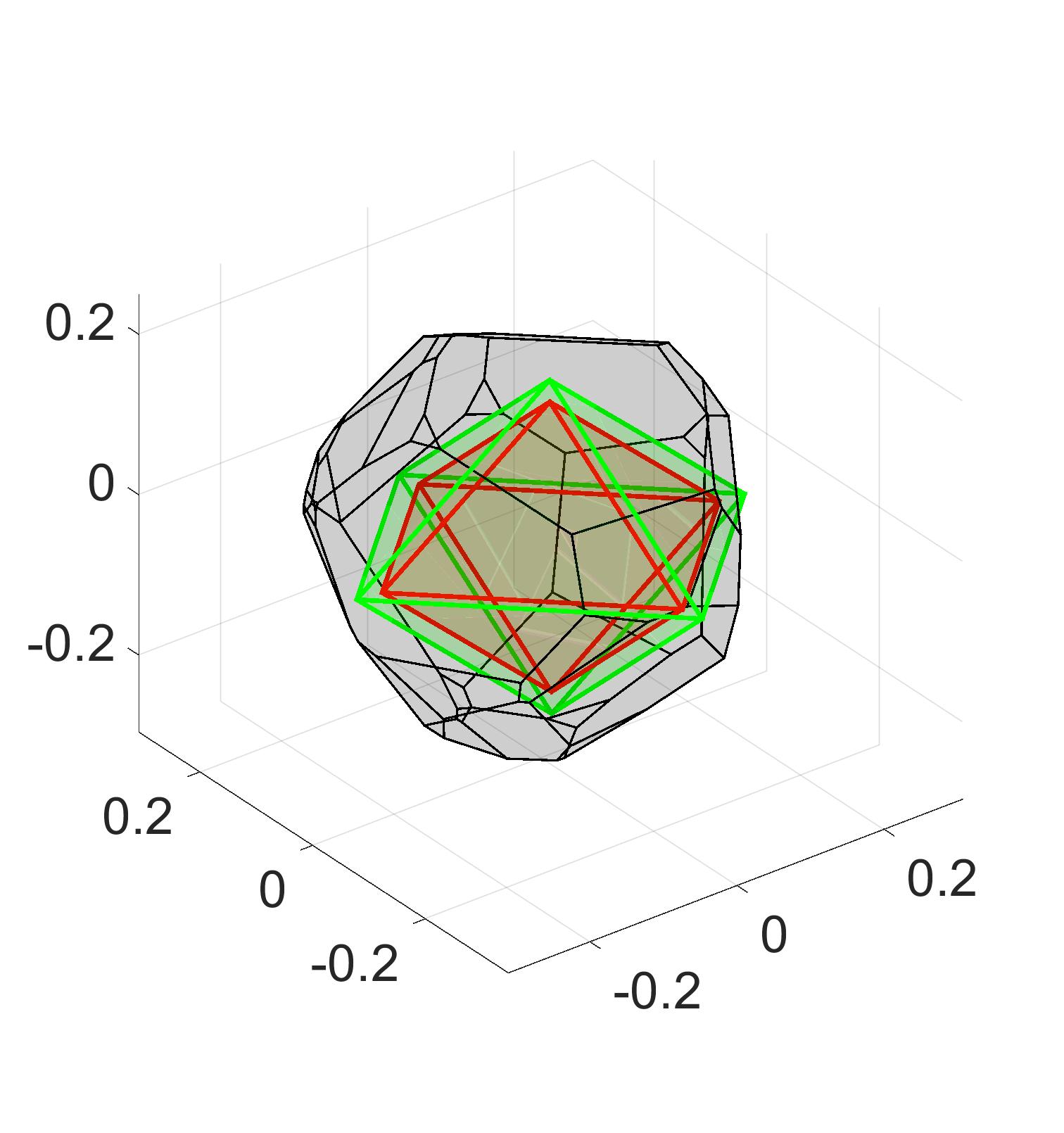}%
\label{f:diam_N1000_16JAN2020}}
\hfil
\subfigure[$\gamma_\infty=1.1890$]{\includegraphics[trim= .5cm .5cm .5cm 2cm, clip=true,width=.49\columnwidth]{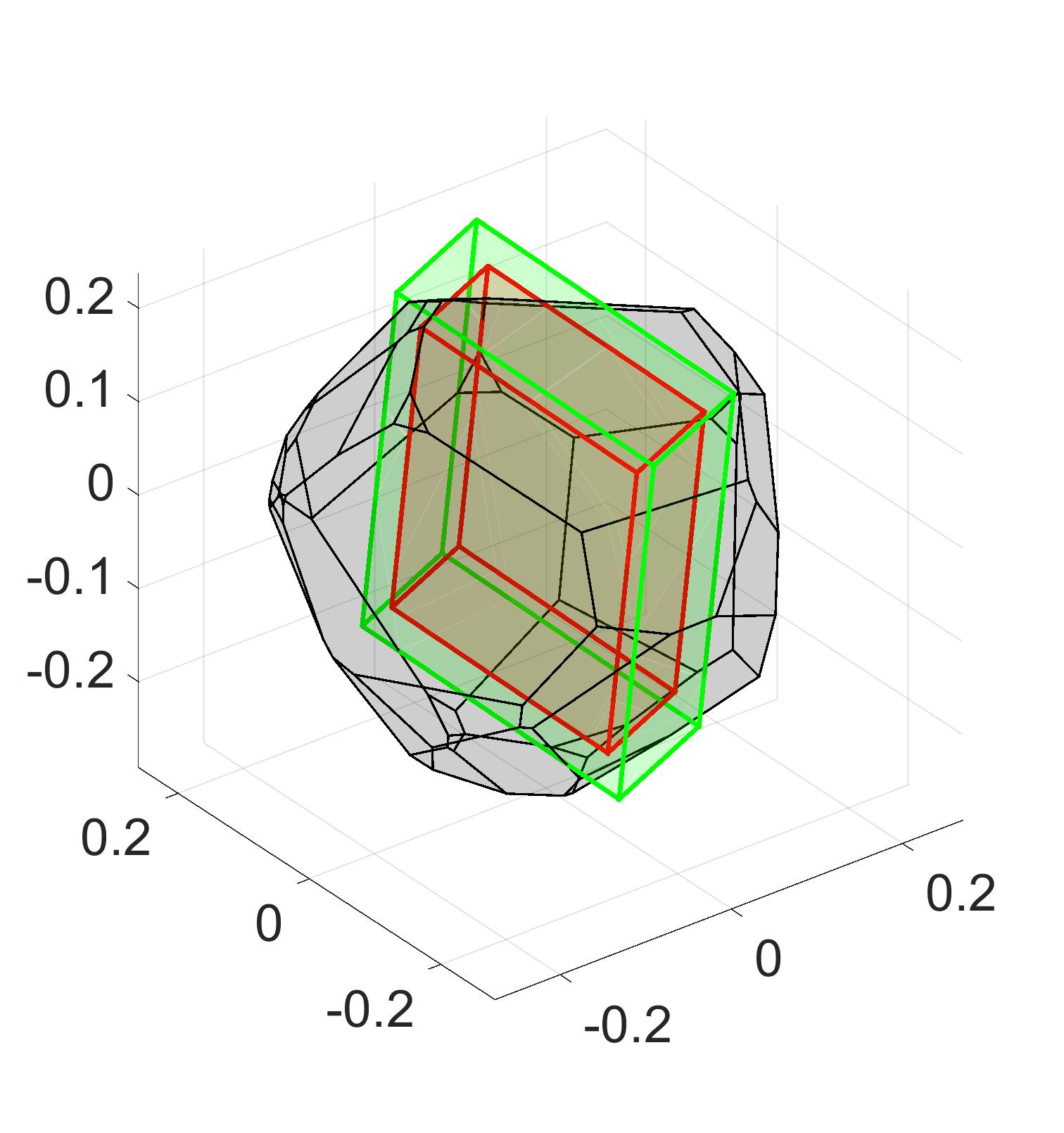}%
\label{f:hyper_N1000_16JAN2020}}
\caption{Scaled $\ell_{1}$-poly (a) and $\ell_{\infty}$-poly (b) obtained starting from $N_D=1,000$.}
\label{f:diam_hyper_1000}
\end{figure}

To better illustrate the proposed approach, and to highlight its main features, we first consider a simple three-dimensional examples ($n_\xi=3$), with scalar uncertain linear inequalities of the form
\[
f(q)^T \xi \le 1
\]
with $f(q)=q_1 q_2$, with $q_1\in\R$ uniformly distributed in the interval $[0.5,1.5]$ and $q_2\in\R^3$ zero-mean Gaussian distribution. Note that, for $n_\xi=3$, the $\ell_1$ and $\ell_{\infty}$-polys have $10$ and $6$ facets, respectively, irrespective to the number of design samples $N_D$ used to preliminary obtain the generic polyhedron $\D$. However, as we will see, the number of constraints $N_D$ employed to design the initial SAS plays a significant role in the final outcome of the procedure. To show this, we performed two different tests, where the number of design samples was set to $N_D=100$ and $N_D=1,000$. The results are shown in Figures \ref{f:diam_hyper_100} and 
\ref{f:diam_hyper_1000} respectively, for both the $\ell_1$ (left) and $\ell_\infty$ (right) cases. Algorithm 1 was applied in all cases with $\varepsilon=0.05$ and $\delta=10^{-6}$, 
leading to $N_\gamma=2,063$ and $r=103$. For allowing a better comparison, the same set of samples where considered for the evaluation of the scaling factor in all examples. These samples lead to $N_\gamma$ random hyper-planes which define a polyhedron represented (in black) in the figures. 

It can be observed that when $N_D$ is small, the ensuing initial $\ell_1$- (resp. $\ell_\infty$-) poly is large, and Algorithm 1 returns a scaling factor $\gamma$ which is less than one (Fig. \ref{f:diam_hyper_100}). Hence, the probabilistic scaling produces a "deflation" of the original set so to guarantee the probabilistic constraints. Vice-versa, for large $N_D$ (Fig. \ref{f:diam_hyper_1000}), the scaling produces an inflation, returning a value of $\gamma$ larger than one.

\begin{figure}[!ht]
\centering
\subfigure[$N_S=10$, $\gamma_S=0.1754$]{\includegraphics[trim= .5cm .5cm .5cm 2cm, clip=true,width=.49\columnwidth]{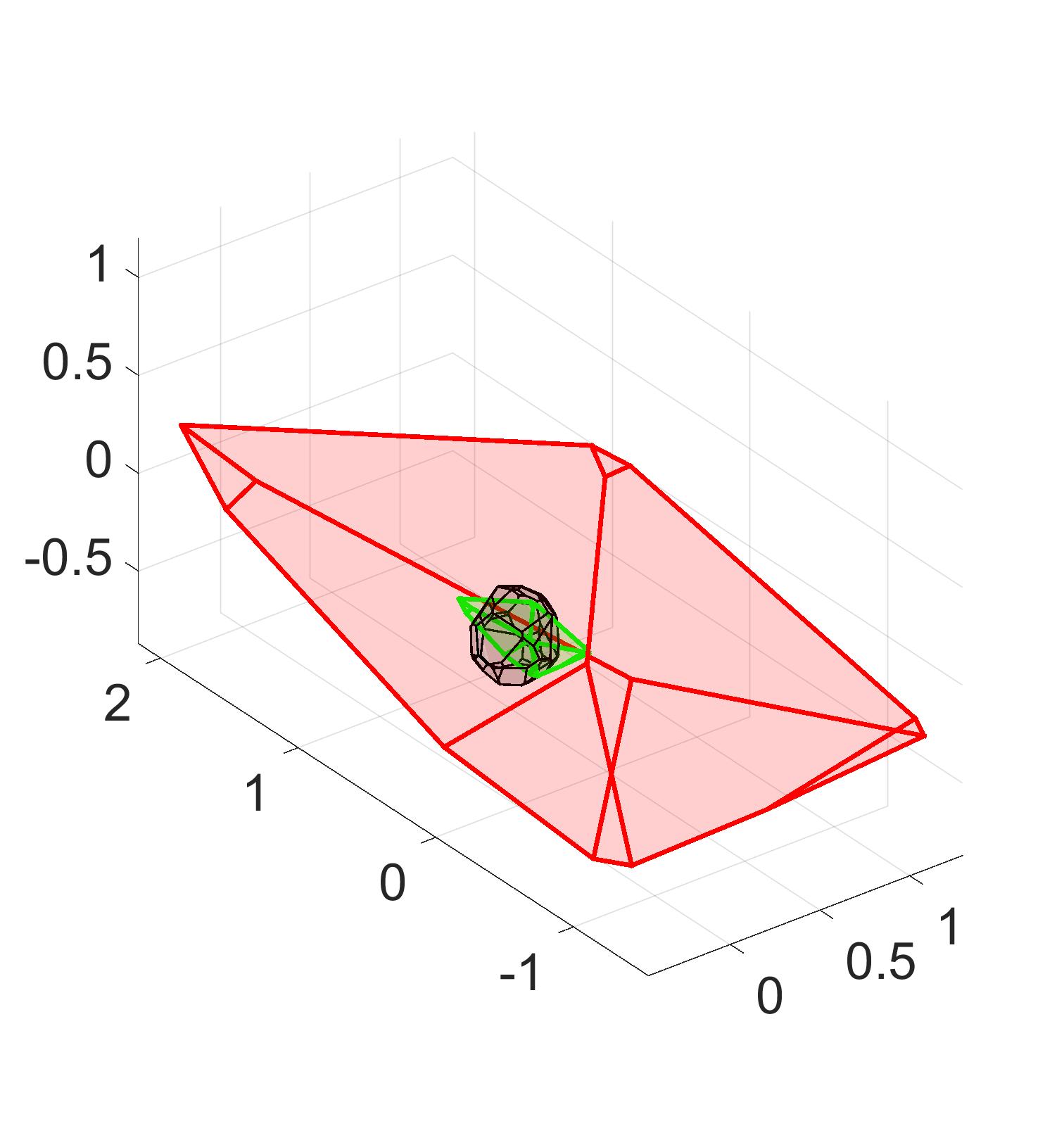}%
\label{f:poly_N10_16JAN2020}}\\
\hfil
\subfigure[$N_S=100$, $\gamma_S=0.6918$]{\includegraphics[trim= .5cm .5cm .5cm 2cm, clip=true,width=.49\columnwidth]{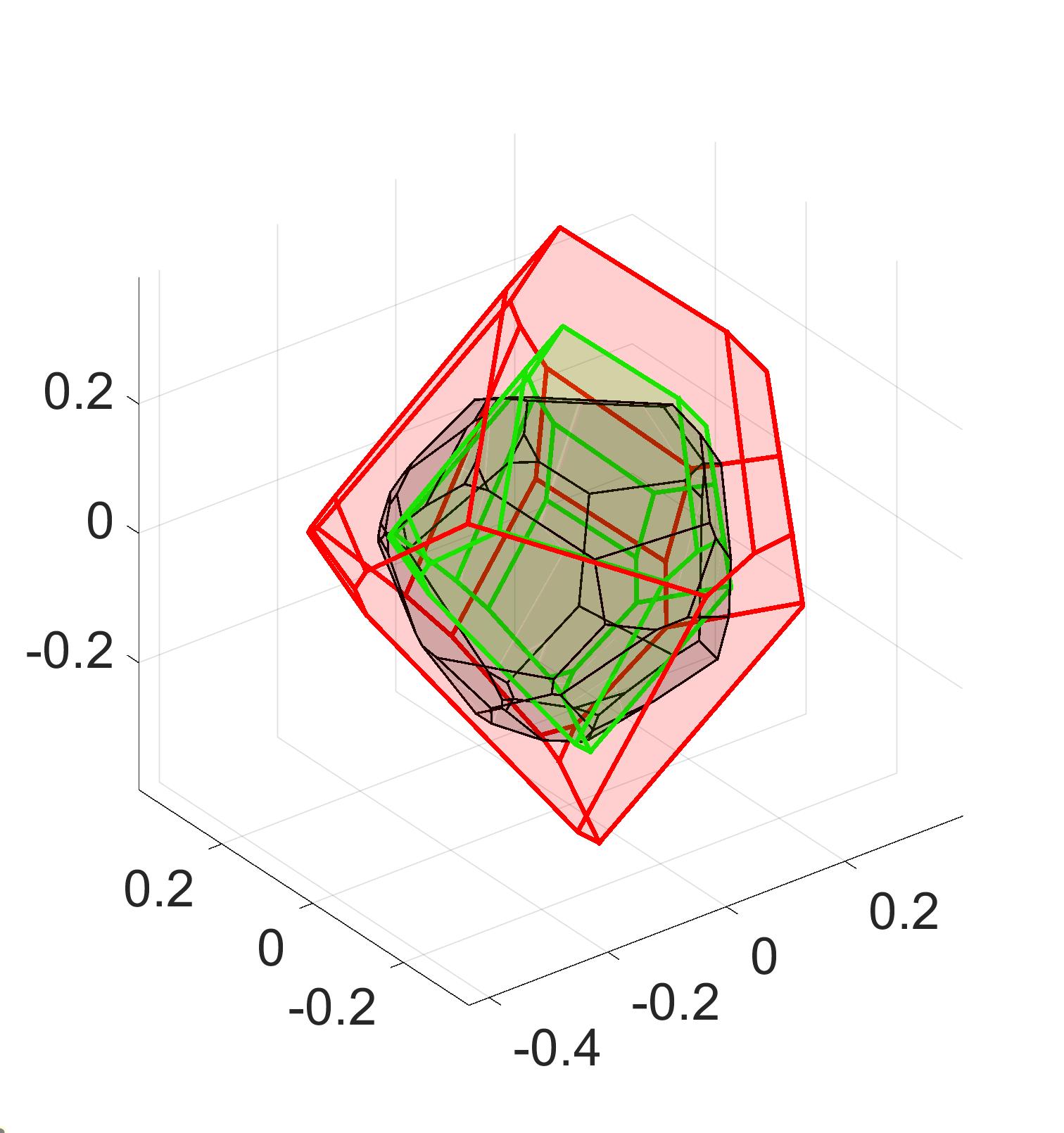}%
\label{f:poly_N100_16JAN2020}}
\hfil
\subfigure[$N_S=1000$, $\gamma_S=0.9459$]{\includegraphics[trim= .5cm .5cm .5cm 2cm, clip=true,width=.49 \columnwidth]{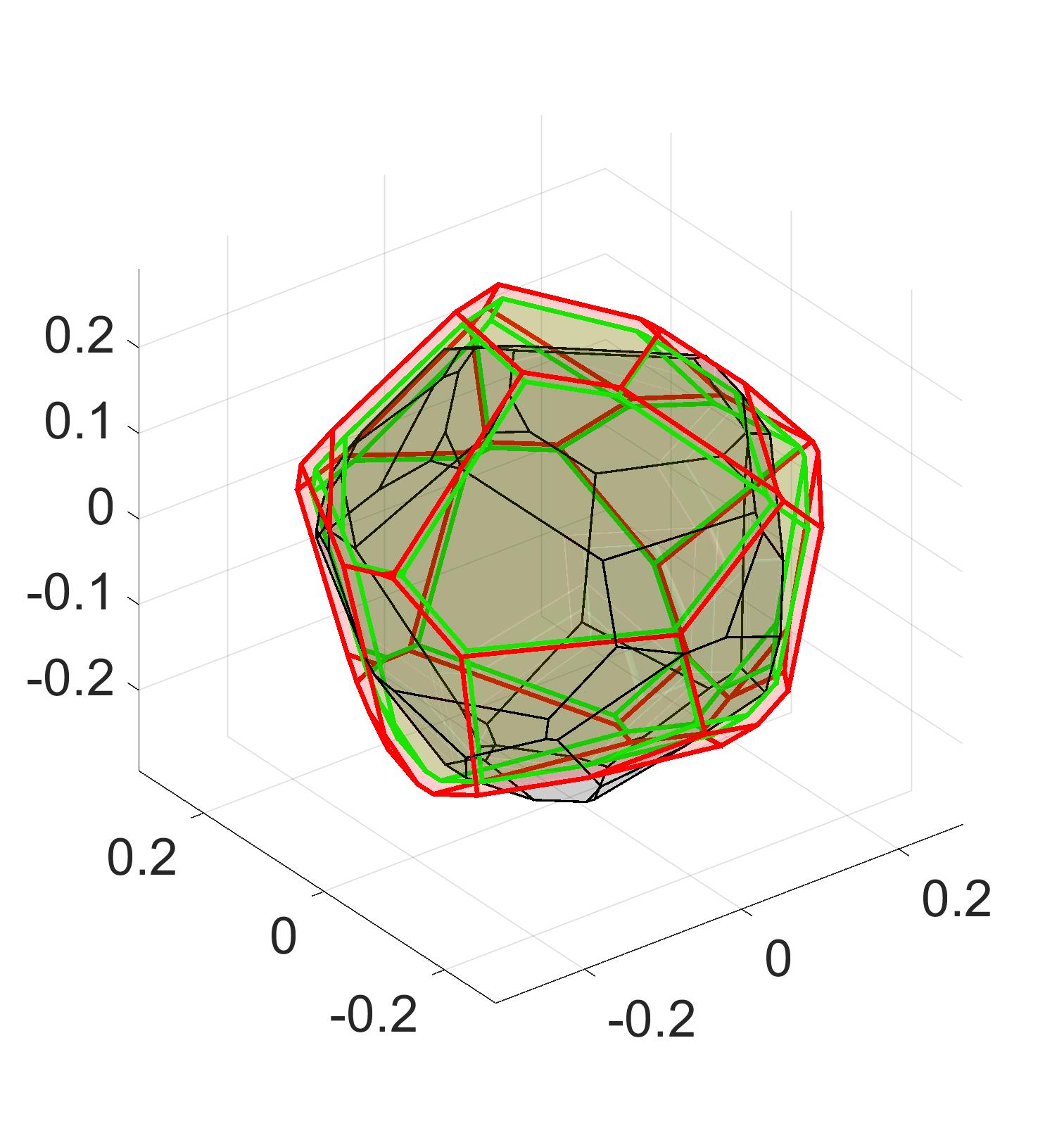}%
\label{f:poly_N1000_16JAN2020}}
\caption{Scaled sampled-polys obtained starting from $N_S=10$ (a), $N_S=100$ (b) and $N_S=1,000$ (c) and for $N_\gamma=2,063$.}
\label{f:poly_16JAN2020}
\end{figure}

Finally, we compared the $\ell_1$- / $\ell_\infty$- polys with the naive approach based on sampled polytope $\Su_S$. Notice that, to allow a fair comparison, we should select a number of hyper-planes comparable with the number of linear inequalities defining $\Su_1$ and $\Su_\infty$. In Fig. \ref{f:poly_N10_16JAN2020}, we represent the initial and final polytopes. Then, we also generated two additional sampled-polys with $N_S=100$ and $N_S=1,000$, i.e. equal to the number of hyper-planes used to generate the design polyhedrons $\D$ for the previous case.
These are depicted in Figs. \ref{f:poly_N100_16JAN2020}-\ref{f:poly_N1000_16JAN2020} while the volumes of the different SASs are reported in Table \ref{tab:volume}.
\begin{table}[!ht]
\begin{center}\caption{Volume of the different SASs considered in Example 1.}
\begin{tabular}{c|cc|cc}
$\Su$ & $N_S$ & $N_D$ & $V$\\
\hline
$\Su_{10}$      & $10$ & $-$   & $0.0091$ \\
$\Su_{100}$      & $100$ & $-$  & $0.0403$ \\
$\Su_{1000}$      & $1000$ & $-$ & $0.0526$ \\
$\Su_1$      & $-$ & $100$  & $0.0176$ \\
$\Su_1$      & $-$ & $1000$ & $0.0258$ \\
$\Su_{\infty}$ & $-$ & $100$  & $0.0131$ \\
$\Su_{\infty}$ & $-$ & $1000$ & $0.0175$\\
\hline 
\end{tabular}
\label{tab:volume}
\end{center}
\end{table}

\section{UAV control over a sloped vineyard}
\label{sec:results}
\begin{figure}[!ht]
\centering
\includegraphics[width=1\columnwidth]{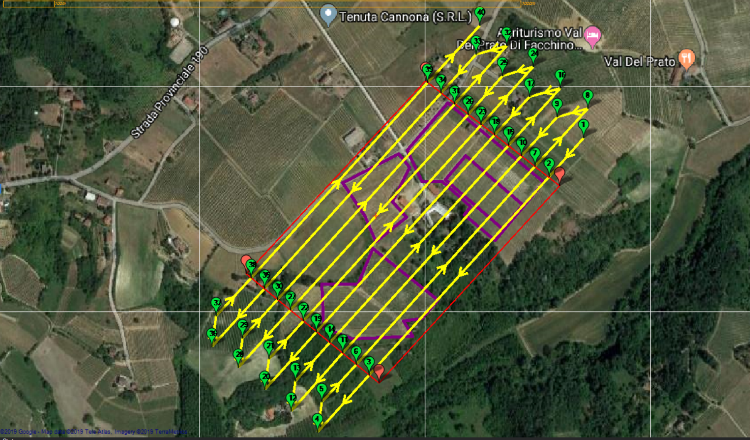}
\caption{Carpeneto vineyard, Piedmont, Italy (credit: Google).}
\label{f:vineyard_WPs1}
\end{figure}

The selected application involves a fixed-wing UAV performing a monitoring mission over a Dolcetto vineyard at Carpeneto, Alessandria, Italy ($44^{\circ}40'55.6''\text{N}, 8^{\circ}37'28.1''\text{E}$). The Mission Planner of ArduPilot open source autopilot has been used to identify a grid pattern with a peculiar path orientation with respect to the grapevine rows, as shown in Fig. \ref{f:vineyard_WPs1}.
The main objective is to provide proper control capabilities to a fixed-wing UAV to guarantee a fixed relative altitude with respect to the terrain of $150$ m while following the desired optimal path defined by the guidance algorithm (described in detail in \cite{mammarella2019waypoint}), maintaining a constant airspeed, i.e. $V_{ref}=12$ m/s. The controllability of the aircraft shall be guaranteed despite the presence of external disturbance due to a fixed-direction wind turbulence, which intensity can randomly vary among $\pm1$ m/s.

For validation purpose, the longitudinal control of the UAV has been provided exploiting both OS-SMPC and the new PS-SMPC approach. In this case study, we have that the state variable are the longitudinal component of the total airspeed in body axes $u$, the angle of attack $\alpha$, the pitch angle $\theta$, the pitch rate $q$, and the altitude $h$. On the other hand, the control variables are represented by the throttle command $\Delta T$ and the elevator deflection $\delta_e$. Hence, we have $n=5$ and $m=2$ while the prediction horizon $T$ has been set equal to $15$. Consequently, setting $\varepsilon=0.05$, $\delta=10^{-6}$, we get $N_{LT}=20,604$ and $N_\gamma=2,063$. On the other hand, the sample complexity selected for generating the $\ell_1$-poly has been set equal to $N_D=100$ obtaining $(n+mT)\cdot N_{D}=3,500$ hyper-planes but only $3(n+mT)+1=107$ linear constraints implemented online.
\begin{figure}[!ht]
\centering
\subfigure[OS-SMPC]{\includegraphics[width=.7\columnwidth]{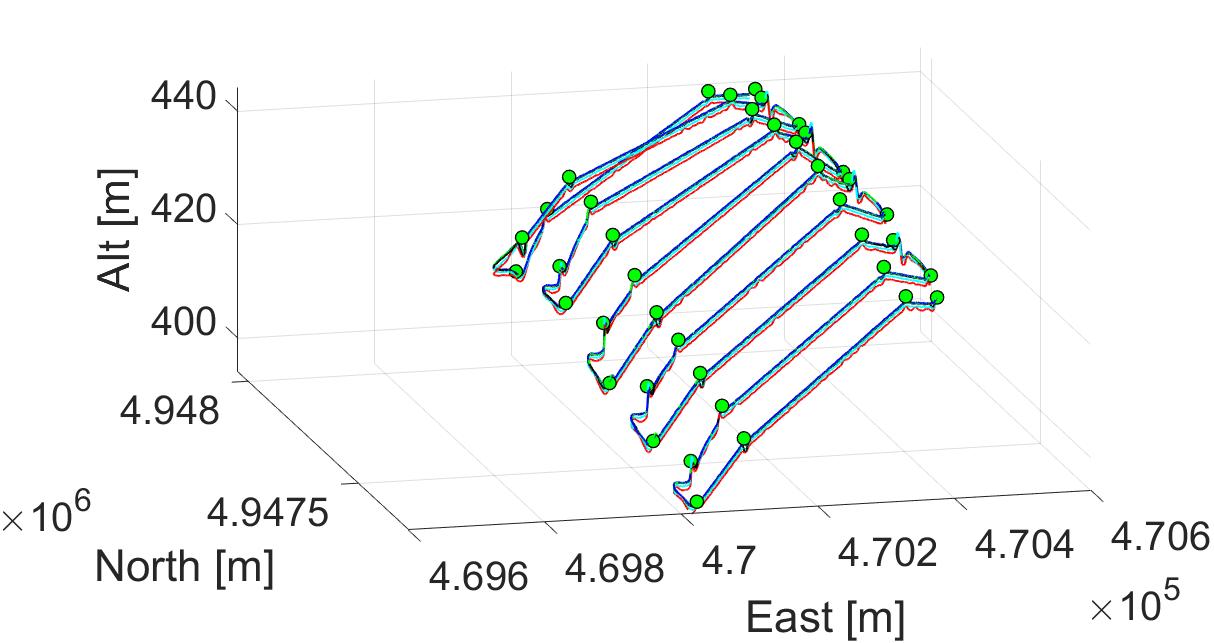}%
\label{f:3D_traj_CCTA_OSSMPC}}
\hfil
\subfigure[PS-SMPC]{\includegraphics[width=.7 \columnwidth]{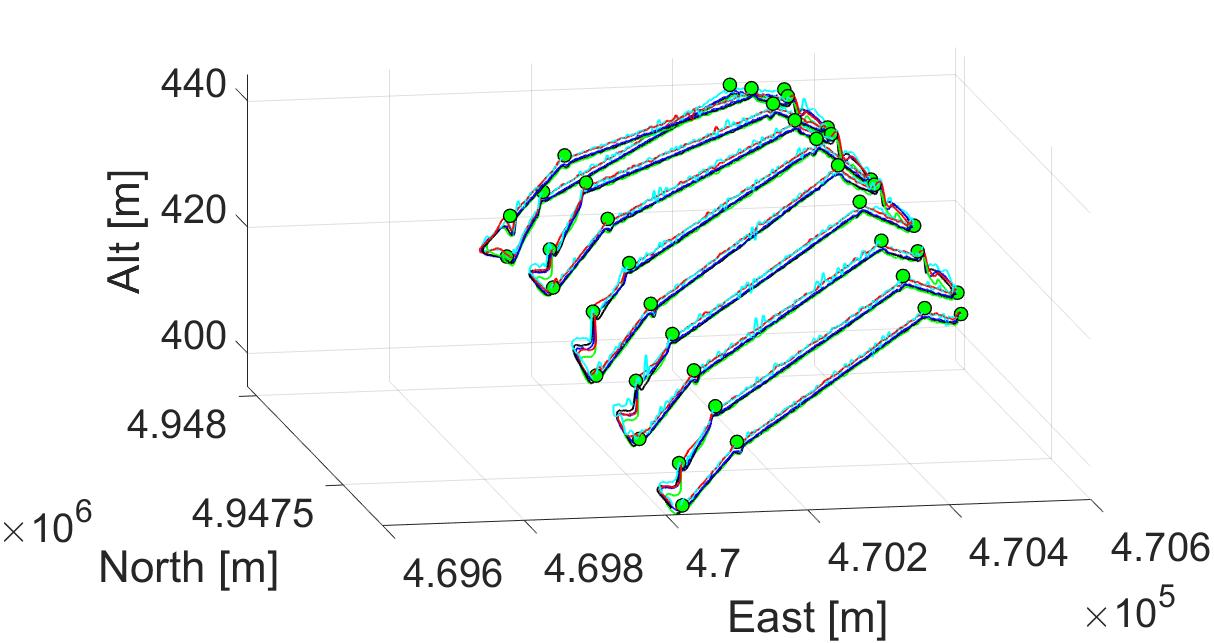}%
\label{f:3D_traj_CCTA_PSSMPC}}
\caption{UAV controlled trajectories obtained running OS-SMPC (a) and PS-SMPC (b) five times each.}
\label{f:3D_traj}
\end{figure}

\begin{figure}[!ht]
\centering
\includegraphics[width=1\columnwidth]{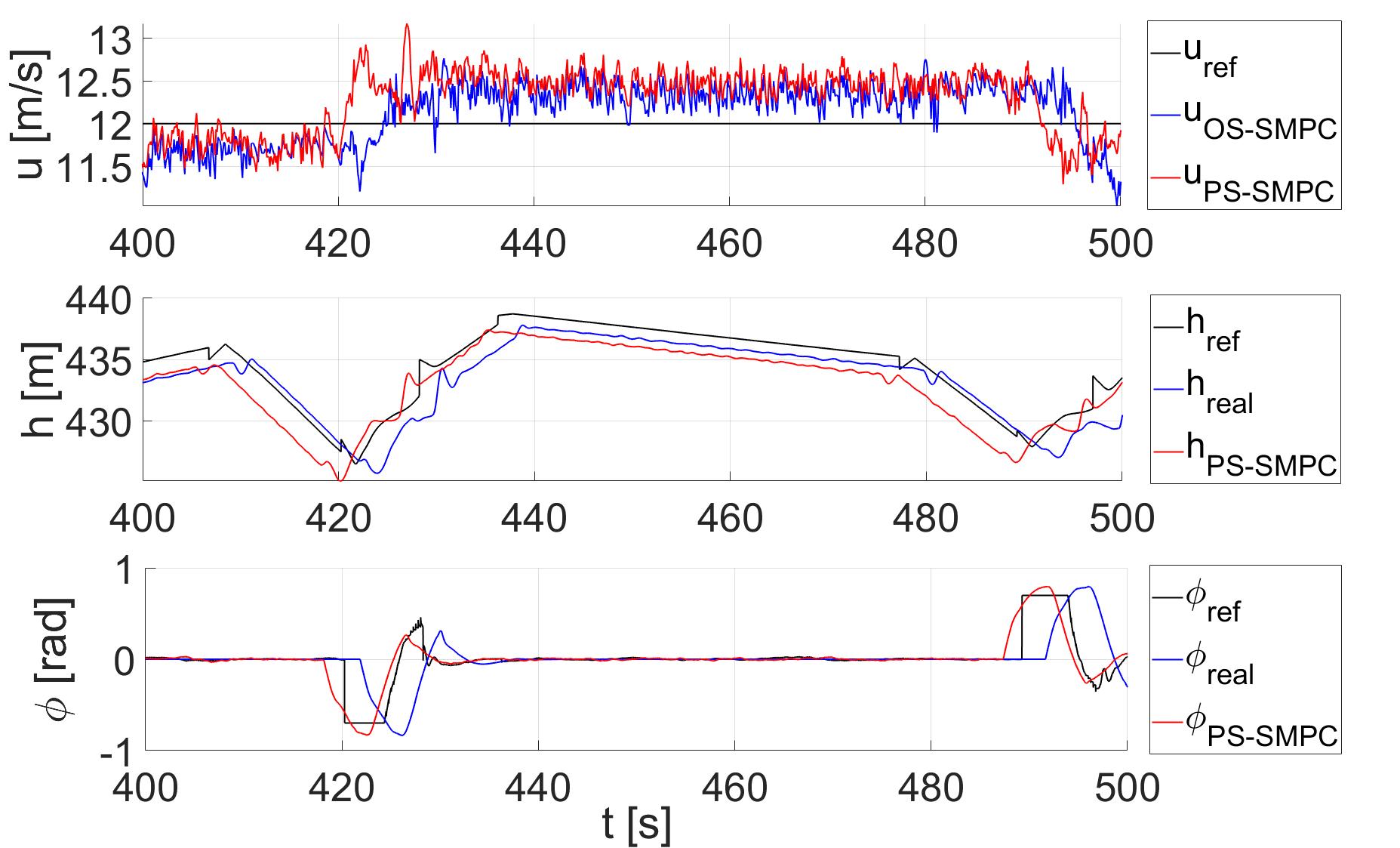}
\caption{Zoom-in on the behavior of controlled state variables, i.e. airspeed $u$, altitude $h$ and roll angle $\phi$, obtained exploiting OS-SMPC (blue lines) and PS-SMPC (red lines) with respect to corresponding reference signals (black lines), i.e. $u_{ref}$, $h_{ref}$ and $\phi_{ref}$.}
\label{f:u_h_phi_CCTA}
\end{figure}

The preliminary results are represented in Fig. \ref{f:3D_traj} as 3D trajectories and in Fig. \ref{f:u_h_phi_CCTA} as controlled states with respect to reference signals. We can notice that both MPC schemes provide acceptable tracking capabilities, despite larger (but still acceptable) oscillations can be observed in \ref{f:3D_traj_CCTA_PSSMPC} when the scaled set is exploited. More interesting results are reported in Tab. \ref{tab:t_comp} in terms of maximum and average values of the computational time required to solve \textit{online} the finite-horizon optimal control problem, evaluated for $5$ different run each. The results show a significant reduction (about 100 times lower) of the computational load when a lower complexity constraint set is employed. This makes the stochastic MPC approach not only effective from a performance viewpoint but also presumably compliant with the computational constraint coming from autopilot hardware.

\begin{table}[!ht]
\begin{center}\caption{Maximum and average computational cost required by OS-SMPC and PS-SMPC approaches for online solving the optimization problem during each run.}
\begin{tabular}{ccccc}
n. & $t_{c_{MAX_{OS}}}$ & $t_{c_{AVG_{OS}}}$ & $t_{c_{MAX_{PS}}}$ & $t_{c_{AVG_{PS}}}$\\
\vspace{0.001cm}\\
\hline
1 & $2.0959$ & $0.4178$ & $0.0966$ & $0.0087$\\

2 & $0.5394$ & $0.3291$ & $0.0215$ & $0.0088$\\

3 & $5.1215$ & $0.4546$ & $0.1065$ & $0.0045$\\

4 & $2.1497$ & $0.5434$ & $0.2628$ & $0.0086$\\

5 & $2.9411$ & $0.5626$ & $0.7221$ & $0.0190$\\
\hline 
\end{tabular}
\label{tab:t_comp}
\end{center}
\end{table}

\section{Conclusions}
\label{sec:concl}
In this paper, we proposed a novel approach which exploits a  probabilistic scaling technique recently proposed by some of the authors to derive a novel Stochastic MPC scheme. The introduced framework exhibits a lower computational complexity, while sharing the appealing probabilistic guarantees of off-line sampling.
\appendix
The proof of to Lemma 1 follows from Proposition 1 in \cite{alamo2019safe}, which guarantees that, for given $r\ge 0$,
$\mathsf{Pr}\{\Su(\gamma) \subseteq \Xe\}$
is guaranteed if the scaling is performed on a number of samples such that
\begin{equation}\label{ineq:N}
N \geq \frac{1}{\varepsilon} \left( r-1+\ln\frac{1}{\delta}+\sqrt{2(r-1)\ln\frac{1}{\delta}}\right).
\end{equation}

Since $r=\lceil \frac{\varepsilon N}{2} \rceil$, we have that $r-1 \leq  \frac{\varepsilon N}{2}$. Thus, inequality \eqref{ineq:N} is satisfied if 
\begin{eqnarray*}
N &\geq& \frac{1}{\varepsilon} \left( \frac{\varepsilon N}{2}+\ln\frac{1}{\delta}+\sqrt{\varepsilon N\ln\frac{1}{\delta}}\right)\\
&=&
 \frac{ N}{2}+\frac{1}{\varepsilon} \ln\frac{1}{\delta}+\sqrt{N\frac{1}{\varepsilon} \ln\frac{1}{\delta}}.
\end{eqnarray*}
Letting%
\footnote{Note that both quantities under square root are positive.}
$\nabla\doteq \sqrt{N}$ and $\alpha\doteq \sqrt{\frac{1}{\varepsilon} \ln\frac{1}{\delta}}$, the above inequality rewrites
$
\nabla^2-2\alpha\nabla -2\alpha^2 \ge 0$,
which has unique positive solution $\nabla\ge (1+\sqrt{3})\alpha$, which rewrites as 
$
N \ge  \frac{(1+\sqrt{3})^2}{\varepsilon} \ln\frac{1}{\delta}$.

The formula in Algorithm 1 follows by observing that $(1+\sqrt{3})^2<7.67$.
\bibliographystyle{IEEEtran}
\bibliography{main}             

\begin{thebibliography}{10}
\providecommand{\url}[1]{#1}
\csname url@samestyle\endcsname
\providecommand{\newblock}{\relax}
\providecommand{\bibinfo}[2]{#2}
\providecommand{\BIBentrySTDinterwordspacing}{\spaceskip=0pt\relax}
\providecommand{\BIBentryALTinterwordstretchfactor}{4}
\providecommand{\BIBentryALTinterwordspacing}{\spaceskip=\fontdimen2\font plus
\BIBentryALTinterwordstretchfactor\fontdimen3\font minus
  \fontdimen4\font\relax}
\providecommand{\BIBforeignlanguage}[2]{{%
\expandafter\ifx\csname l@#1\endcsname\relax
\typeout{** WARNING: IEEEtran.bst: No hyphenation pattern has been}%
\typeout{** loaded for the language `#1'. Using the pattern for}%
\typeout{** the default language instead.}%
\else
\language=\csname l@#1\endcsname
\fi
#2}}
\providecommand{\BIBdecl}{\relax}
\BIBdecl

\bibitem{farina2016stochastic}
M.~Farina, L.~Giulioni, and R.~Scattolini, ``Stochastic linear model predictive
  control with chance constraints--a review,'' \emph{Journal of Process
  Control}, vol.~44, pp. 53--67, 2016.

\bibitem{tempo2012randomized}
R.~Tempo, G.~Calafiore, and F.~Dabbene, \emph{Randomized algorithms for
  analysis and control of uncertain systems: with applications}.\hskip 1em plus
  0.5em minus 0.4em\relax Springer Science \& Business Media, 2012.

\bibitem{calafiore2006scenario}
G.~C. Calafiore and M.~C. Campi, ``The scenario approach to robust control
  design,'' \emph{IEEE Transactions on Automatic Control}, vol.~51, no.~5, pp.
  742--753, 2006.

\bibitem{schildbach2014scenario}
G.~Schildbach, L.~Fagiano, C.~Frei, and M.~Morari, ``The scenario approach for
  stochastic model predictive control with bounds on closed-loop constraint
  violations,'' \emph{Automatica}, vol.~50, no.~12, pp. 3009--3018, 2014.

\bibitem{grosso2017stochastic}
J.~M. Grosso, P.~Velarde, C.~Ocampo-Martinez, J.~M. Maestre, and V.~Puig,
  ``Stochastic model predictive control approaches applied to drinking water
  networks,'' \emph{Optimal Control Applications and Methods}, vol.~38, no.~4,
  pp. 541--558, 2017.

\bibitem{nasir2015randomised}
H.~A. Nasir, A.~Car{\`e}, and E.~Weyer, ``A randomised approach to flood
  control using value-at-risk,'' in \emph{2015 54th IEEE Conference on Decision
  and Control (CDC)}.\hskip 1em plus 0.5em minus 0.4em\relax IEEE, 2015, pp.
  3939--3944.

\bibitem{van2006stochastic}
D.~Van~Hessem and O.~Bosgra, ``Stochastic closed-loop model predictive control
  of continuous nonlinear chemical processes,'' \emph{Journal of Process
  Control}, vol.~16, no.~3, pp. 225--241, 2006.

\bibitem{vignali2017energy}
R.~M. Vignali, F.~Borghesan, L.~Piroddi, M.~Strelec, and M.~Prandini, ``Energy
  management of a building cooling system with thermal storage: An approximate
  dynamic programming solution,'' \emph{IEEE Transactions on Automation Science
  and Engineering}, vol.~14, no.~2, pp. 619--633, 2017.

\bibitem{matthias1}
M.~Lorenzen, F.~Dabbene, R.~Tempo, and F.~Allg{\"o}wer, ``Stochastic {MPC} with
  offline uncertainty sampling,'' \emph{Automatica}, vol.~81, no.~1, pp.
  176--183, 2017.

\bibitem{Mammarella:18:Control:Systems:Technology}
M.~{Mammarella}, M.~{Lorenzen}, E.~{Capello}, H.~{Park}, F.~{Dabbene},
  G.~{Guglieri}, M.~{Romano}, and F.~{Allg\"ower}, ``An offline-sampling {SMPC}
  framework with application to autonomous space maneuvers,'' \emph{IEEE
  Transactions on Control Systems Technology}, pp. 1--15, 2018.

\bibitem{kamel}
M.~Kamel, T.~Stastny, K.~Alexis, and R.~Siegwart, ``Model predictive control
  for trajectory tracking of unmanned aerial vehicles using robot operating
  system,'' in \emph{Robot Operating System (ROS)}.\hskip 1em plus 0.5em minus
  0.4em\relax Springer, 2017, pp. 3--39.

\bibitem{Alexis2016}
K.~Alexis, C.~Papachristos, R.~Siegwart, and A.~Tzes, ``Robust model predictive
  flight control of unmanned rotorcrafts,'' \emph{Journal of Intelligent \&
  Robotic Systems}, vol.~81, no. 3-4, pp. 443--469, 2016.

\bibitem{stastny}
T.~J. Stastny, A.~Dash, and R.~Siegwart, ``Nonlinear mpc for fixed-wing {UAV}
  trajectory tracking: Implementation and flight experiments,'' in \emph{AIAA
  Guidance, Navigation, and Control Conference}, 2017, p. 1512.

\bibitem{michel}
N.~Michel, S.~Bertrand, G.~Valmorbida, S.~Olaru, and D.~Dumur, ``Design and
  parameter tuning of a robust model predictive controller for {UAV}s,'' in
  \emph{2017 20th IFAC World Congress}, 2017.

\bibitem{mammarella2018sample}
M.~Mammarella, E.~Capello, F.~Dabbene, and G.~Guglieri, ``Sample-based {SMPC}
  for tracking control of fixed-wing {UAV},'' \emph{IEEE Control Systems
  Letters}, vol.~2, no.~4, pp. 611--616, 2018.

\bibitem{alamo2019safe}
T.~Alamo, V.~Mirasierra, F.~Dabbene, and M.~Lorenzen, ``Safe approximations of
  chance constrained sets by probabilistic scaling,'' in \emph{2019 18th
  European Control Conference (ECC)}.\hskip 1em plus 0.5em minus 0.4em\relax
  IEEE, 2019, pp. 1380--1385.

\bibitem{dabbene2010complexity}
F.~Dabbene, C.~Lagoa, and P.~Shcherbakov, ``On the complexity of randomized
  approximations of nonconvex sets,'' in \emph{2010 IEEE International
  Symposium on Computer-Aided Control System Design}.\hskip 1em plus 0.5em
  minus 0.4em\relax IEEE, 2010, pp. 1564--1569.

\bibitem{sylvester2018agriculture}
G.~Sylvester, G.~Rambaldi, D.~Guerin, A.~Wisniewski, N.~Khan, J.~Veale, and
  M.~Xiao, ``E-agriculture in action-drones for agriculture. food and
  agriculture organization of the united nations and international
  telecommunication union,'' \emph{Bangkok: Food and Agriculture Organization
  of the United Nations and International Telecommunication Union. Retrieved
  July}, vol.~19, 2018.

\bibitem{matthias2}
M.~Lorenzen, F.~Dabbene, R.~Tempo, and F.~Allg{\"o}wer, ``Constraint-tightening
  and stability in stochastic model predictive control,'' \emph{IEEE
  Transactions on Automatic Control}, vol.~62, no.~7, pp. 3165--3177, 2017.

\bibitem{vidyasagar}
M.~Vidyasagar, \emph{Learning and Generalisation: with Applications to Neural
  Networks}.\hskip 1em plus 0.5em minus 0.4em\relax Springer Science \&
  Business Media, 2013.

\bibitem{alamo2009randomized}
T.~Alamo, R.~Tempo, and E.~F. Camacho, ``Randomized strategies for
  probabilistic solutions of uncertain feasibility and optimization problems,''
  \emph{IEEE Transactions on Automatic Control}, vol.~54, no.~11, pp.
  2545--2559, 2009.

\bibitem{yan2018stochastic}
S.~Yan, P.~Goulart, and M.~Cannon, ``Stochastic model predictive control with
  discounted probabilistic constraints,'' in \emph{2018 European Control
  Conference (ECC)}.\hskip 1em plus 0.5em minus 0.4em\relax IEEE, 2018, pp.
  1003--1008.

\bibitem{hewing2018stochastic}
L.~Hewing and M.~N. Zeilinger, ``Stochastic model predictive control for linear
  systems using probabilistic reachable sets,'' in \emph{2018 IEEE Conference
  on Decision and Control (CDC)}, 2018, pp. 5182--5188.

\bibitem{kaibel2003some}
V.~Kaibel and M.~E. Pfetsch, ``Some algorithmic problems in polytope theory,''
  in \emph{Algebra, geometry and software systems}.\hskip 1em plus 0.5em minus
  0.4em\relax Springer, 2003, pp. 23--47.

\bibitem{mammarella2019waypoint}
M.~Mammarella, G.~Ristorto, E.~Capello, N.~Bloise, G.~Guglieri, and F.~Dabbene,
  ``Waypoint tracking via tube-based robust model predictive control for crop
  monitoring with fixed-wing {UAV}s,'' in \emph{2019 IEEE International
  Workshop on Metrology for Agriculture and Forestry (MetroAgriFor)}.\hskip 1em
  plus 0.5em minus 0.4em\relax IEEE, 2019, pp. 19--24.

\end{thebibliography}

\end{document}